\documentclass[twocolumn,aps,showpacs,floatfix,prc]{revtex4}
\usepackage[dvips]{epsfig}
\usepackage{subeqn}

\begin{document}

\title{ Proton radioactivity half lives with Skyrme interactions }

\author{ T. R. Routray$^1$, Abhishek Mishra$^2$, S. K. Tripathy$^{1,3}$, B. Behera$^1$ and D. N. Basu$^2$ }

\affiliation{ $^1$School of Physics, Sambalpur University- 768019, Orissa, India }
\affiliation{ $^2$Variable Energy Cyclotron Centre, 1/AF Bidhan Nagar, Kolkata 700 064, India }
\affiliation{$^3$Govt. Engg. College, Bhawanipatna, Kalhandi, Orissa, INDIA}

\email[E-mail 1: ]{trr1@rediffmail.com}
\email[E-mail 2: ]{abhishek.mishra@vecc.gov.in}
\email[E-mail 3: ]{tripathy_sunil@rediffmail.com}
\email[E-mail 4: ]{dnb@vecc.gov.in}

\date{\today }

\begin{abstract}

    The potential barrier impeding the spontaneous emission of protons in the proton radioactive nuclei is calculated as the sum of nuclear, Coulomb and centrifugal contributions. The nuclear part of the proton-nucleus interaction potential is obtained in the energy density formalism using Skyrme effective interaction that results into a simple algebraic expression. The half-lives of the proton emitters are calculated for the different Skyrme sets within the improved WKB framework. The results are found to be in reasonable agreement with the earlier results obtained for more complicated calculations involving finite range interactions. 
\vspace{0.25cm}


\end{abstract}


\maketitle

\section{Introduction}

    The phenomenon of proton radioactivity is much recent compared to $\alpha$ radioactivity and has been  possible with the fusion-evaporation reactions between stable nuclei and, subsequently, with the advent of radioactive ion beam facilities. The neutron deficient nuclei lying beyond the proton drip-line have positive Q-values for proton emissions and are spontaneous proton emitters. This limits the possibilities of detection of even more exotic nuclei on the proton rich side of the $\beta$-stability valley because of the experimental limitations that the proton emitters can be detected only if they have a half-life longer than $\sim$ 1 $\mu$ sec. So far proton radioactivity has been identified for different isotopes of Sb, Tm, Lu, Ta, Ir, Au and Bi (spherical proton emitters) and of I, Cs, La, Pr, Eu, Tb and Ho (deformed proton emitters) and the only three elements which are missing between Z=51 and Z=83, are Te, Pm and Hg with Z=52, Z=61 and Z=80, respectively.    

    The proton radioactivity has been investigated and half-lives have been calculated in different theoretical models \cite{BMP92,Ab97,BA05,BCS05,De06,MG07,BCS08,YENI11,Fe11}. On the basis of the calculations of proton decay half lives, the theoretical investigations can broadly be divided into two groups. One deals with the quantum mechanical tunneling of single particle resonance through the nuclear mean field \cite{Ab97,Kh07,Fe11}. In these calculations, the mean field potential is calculated self consistently either using covariant density functional theory (CDFT) in the relativistic mean field model (RMF) \cite{Fe11} or from the non-relativistic Hartree-Fock (HF) calculation using phenomenological effective interactions \cite{Kh07}. This procedure of handling the proton radioactivity phenomenon is more fundamental as it deals with the spectroscopic aspect of the emitted proton directly and also has the flexibility to account for the deformation of the emitting nucleus. The other method of calculating proton radioactivity half-lives uses the semiclassical WKB tunneling through the potential barrier. In these calculations the barrier impeding the emission of proton is obtained and the penetration probability is then calculated using WKB approximation. Both the procedures are found to be equally competent \cite{Ab97} so far as the prediction of the proton emission half-lives are concerned. The various ways of constructing the potential barrier in the latter case include phenomenological parametrizations \cite{Gu99}, from fusion reaction studies \cite{BA05}, semiclassical considerations based on liquid drop model and the proximity force \cite{Do09,Do10,Zh09,Zh10} and folding the nucleon-nucleon (N-N) effective interaction over the density distribution of the daughter nucleus \cite{BCS05,MG07,BCS08,YENI11}. The JLM \cite{MG07}, DDM3Y \cite{BCS08} and YENI \cite{YENI11} N-N effective interactions are used in the folding model calculations.
        
    Skyrme interaction \cite{BV72,BV75,BV81} has an important status in the finite nucleus calculations in nuclear research. More than 110 Skyrme sets have been constructed so far for different purposes but all the sets have the common feature of explaining the ground state properties of nuclei over the periodic table and saturation conditions in symmetric nuclear matter (SNM). Skyrme sets constructed in late 90's, particularly the construction of SLy-sets \cite{Ch97,Ch98,St03} and other Skyrme sets developed thereafter have the additional feature that the Skyrme parameters are constrained for application to nuclear matter under extreme conditions. Stone et al. \cite{St03} have examined 87 numbers of Skyrme sets on the basis of various constraints and have found 27 sets qualifying the tests for application to neutron rich dense matter. Proton radioactivity half-lives of the proton emitters for the SkP Skyrme set has been calculated \cite{Kh07} using both the methods, by semiclassical WKB tunneling and also by the quantum mechanical tunneling of the single particle resonance. In the present work, our objective is to calculate the half-lives of the proton emitters for different Skyrme sets in the framework of semiclassical WKB method. The nuclear part of the proton-nucleus (p-N) interaction potential for the Skyrme force is obtained analytically in the energy density formalism where the energy dependence is appropriately taken into account.         
    
    The real part of the p-N interaction potential is calculated for Skyrme effective interaction in section-II. The procedure for calculation of the half-lives of the spontaneous proton emitters is described in this section. In section-III, the results of half-lives for the proton emitters are given for different Skyrme sets. Section-IV contains a brief discussion of the results obtained and conclusion.     
                
\section{Theoretical formalism}

    The energy density, $H(\rho_n,\rho_p)$, of a nucleus is given by

\begin{equation}
 H(\rho_n,\rho_p) =  \frac{\hbar^2}{2m}(\tau_n + \tau_p) +  V
\label{seqn1}
\end{equation}

\noindent  
where $\rho_{n(p)}(\vec{r})$ and $\tau_{n(p)}(\vec{r})$ are the density and the kinetic energy density of the neutron (proton) at position $\vec{r}$ and $m$ is the reduced nucleonic mass for the proton-daughter nucleus system. The first term on the right side of Eq.(1) is the kinetic part of the energy density and the second term is the interaction part which for Skyrme interaction can be expressed as (for details see Refs.\cite{Ch97,Ch98}),

\begin{eqnarray}
 V =&& A(\rho_n,\rho_p) +  B_1\rho \tau  +  B_2(\rho_p \tau_n + \rho_n \tau_p) \nonumber \\
      +&& C [ (\nabla\rho_n)^2 + (\nabla\rho_p)^2 ] + D (\nabla\rho_n)(\nabla\rho_p), 
\label{seqn2}
\end{eqnarray}
\noindent  
with $\rho(\vec{r})=\rho_n(\vec{r})+\rho_p(\vec{r})$ and $\tau(\vec{r})=\tau_n(\vec{r})+\tau_p(\vec{r})$ being the total nucleonic density and total kinetic energy density, respectively, at position $\vec{r}$ and 

\begin{subequations}
\label{allequations}
\begin{eqnarray}
 A(\rho_n,\rho_p) = \frac{t_0}{4} [(1-x_0)(\rho_n^2+\rho_p^2)+(4+2x_0)\rho_n \rho_p]  \nonumber \\ 
 +\frac{t_3}{24}\rho^\gamma [(1-x_3) (\rho_n^2+\rho_p^2) + (4+2x_3)\rho_n\rho_p],   \label{equationa} \\
 B_1 = \frac{1}{8}[t_1(1-x_1)+3t_2(1+x_2)],    \label{equationb}  \\
 B_2 = \frac{1}{8}[t_1(1+2x_1)-t_2(1+2x_2)],   \label{equationc}  \\
 C = \frac{3}{32}[ t_1(1-x_1)-t_2(1+x_2)],     \label{equationd}  \\
 D = \frac{1}{16}[3 t_1(2+x_1)-t_2(2+x_2)].  \label{equatione}
\label{seqn3}
\end{eqnarray}
\end{subequations}
\noindent  

    The real part of the p-N potential is the functional derivative of energy density $H(\rho_n,\rho_p)$ with respect to proton, i.e., $\frac{\partial H}{\partial [f]_p}$. Further the non-local Skyrme Hartree-Fock potential, {\it i.e.} specified by a nucleon effective mass, can be taken care by an equivalent energy dependent local potential \cite{Do72}. Considering this energy dependence of the real part of the nucleon-nucleus potential, the p-N potential $U_N^p(r)$ can now be given by  

\begin{eqnarray}
 U_N^p(r) = \Big[ 1-\Big(\frac{m^*(r)}{m}\Big)_p \Big] (E_{CM}-U_{Coul})   \nonumber  \\
 + \Big(\frac{m^*(r)}{m}\Big)_p \Big[ \frac{\partial A(\rho_n,\rho_p)}{\partial \rho_p} + B_1 \tau + B_2 \tau_n - 2 C \nabla^2 \rho_p  \nonumber  \\
- D \nabla^2 \rho_n + \frac{1}{2} \Big(\frac{d^2}{dr^2}\frac{\hbar^2}{2m^*_p(r)}\Big) - \frac{m^*_p(r)}{2\hbar^2}\Big(\frac{d}{dr}\frac{\hbar^2}{2m^*_p(r)}\Big)^2 \Big] 
\label{seqn4}
\end{eqnarray}
\noindent
where, the functional $A(\rho_n,\rho_p)$ and the functions $B_1$, $B_2$, $C$ and $D$ are given in Eqs.(3.a-e), $E_{CM}$ is the center of mass energy of the proton-daughter nucleus system, $U_{Coul}$ is the Coulomb interaction energy of the proton and $(\frac{m^*(r)}{m})_p$ is the proton effective mass which is given by

\begin{equation}
 (\frac{m^*(r)}{m})_p = \Big[ 1+\frac{2m}{\hbar^2} (B_1 \rho + B_2 \rho_n) \Big]^{-1}.
\label{seqn5}
\vspace{0.0cm}
\end{equation}
\noindent  
In obtaining Eq.(4) we have used the fact that the wave number, $k$, of the proton at a distance $\vec{r}$ from the center of the nucleus taken as the origin is

\begin{equation}
 k(\vec{r}) = \Big[\frac{2m}{\hbar^2} \Big\{ E_{CM}-U_N^p(r)-U_{Coul} \Big\} \Big]^{1/2}.
\label{seqn6}
\vspace{0.0cm}
\end{equation}
\noindent
The total potential, $U^p(r)$, experienced by the proton is the sum of the nuclear potential $U_N^p(r)$, Coulomb potential $U_{Coul}$ and the centrifugal potential $\frac{\hbar^2}{2m}[l(l+1)]$ with $l\hbar$ being the orbital angular momentum carried away by the emitted proton. The Coulomb potential $U_{Coul}$ comprises of the direct ($U^d_{Coul}$) and the exchange ($U^{ex}_{Coul}$) parts, $U_{Coul} = U^d_{Coul} + U^{ex}_{Coul}$, given as

\begin{subequations}
\label{allequations}
\begin{eqnarray}
U^d_{Coul}(r) = 4\pi e^2[\frac{1}{r}\int_0^r r'^2\rho_p(r') dr' \nonumber \\
              + \int_r^\infty r' \rho_p(r') dr'] \label{equationa} \\
U^{ex}_{Coul}(r) = -e^2(\frac{3}{\pi})^\frac{1}{3} \rho_p^\frac{1}{3}(r) \label{equationb} 
\label{seqn7}
\vspace{0.0cm}
\end{eqnarray}
\end{subequations}
\noindent
where $e$ is the electronic charge. The p-N nuclear potential $U_N^p(r)$ for the Skyrme interaction as obtained from Eq.(4) using Eqs.(3.a-e) is given by

\begin{eqnarray}
 &&U_N^p(r) = \Big[1-\Big(\frac{m^*(r)}{m}\Big)_p\Big] (E_{CM}-U_{Coul}) \nonumber  \\
 &&+ \Big(\frac{m^*(r)}{m}\Big)_p \Big[ \frac{t_0}{2} [ (1-x_0)\rho_p+(2+x_0)\rho_n] \nonumber \\ 
 &&+ \frac{t_3}{12}\Big( \gamma [(1-x_3) \frac{\rho_p^2+\rho_n^2}{2} + (2+x_3)\rho_p\rho_n] \nonumber \\
 &&+\rho [(1-x_3)\rho_p+(2+x_3)\rho_n] \Big) \rho^{\gamma-1} \nonumber \\
 &&+\frac{1}{8}[t_1(2+x_1)+t_2(2+x_2)]\tau_n \nonumber \\
 &&+ \frac{1}{8}[t_1(1-x_1)+3t_2(1+x_2)]\tau_p \nonumber \\
 &&-\frac{3}{16}[t_1(1-x_1)-t_2(1+x_2)](\nabla^2\rho_p) \nonumber \\
  &&-\frac{1}{16}[3t_1(2+x_1)-t_2(2+x_2)](\nabla^2\rho_n) \nonumber \\
   &&+\frac{1}{16}\Big\{[t_1(2+x_1)+t_2(2+x_2)]\frac{d^2\rho_n}{dr^2} \nonumber \\
 &&+ [t_1(1-x_1)+3t_2(1+x_2)]\frac{d^2\rho_p}{dr^2}\Big\} \nonumber \\ 
 &&-\frac{m^*_p(r)}{128\hbar^2}\Big\{[t_1(2+x_1)+t_2(2+x_2)]\frac{d\rho_n}{dr} \nonumber \\
 &&+ [t_1(1-x_1)+3t_2(1+x_2)]\frac{d\rho_p}{dr}\Big\}^2 \Big].
\label{seqn8}
\end{eqnarray}
\noindent

    The potential barrier is obtained from $U^p(r)$ and the center of mass energy $E_{CM}=Q$ determines the turning points $R_a$ and $R_b$. The penetration probability, $P$, is now calculated from the improved WKB formula \cite{Ke35},

\begin{equation}
 P= \Big[ 1+\exp \Big\{ \frac{2}{\hbar} \int_{R_a}^{R_b} {[2\mu (U^p(r) - E_v - Q)]}^{1/2} dr \Big\} \Big]^{-1},
\label{seqn9}
\vspace{0.0cm}
\end{equation}
\noindent
where $\mu = mM_d/M_A$  is the reduced mass,  $M_d$ and $M_A$  being the masses of the daughter nucleus and the parent nucleus respectively and $E_v$ is the energy with which the proton hits the barrier. The decay constant $\lambda$ and half life $T_{1/2}$ are now given by    

\begin{equation}
 \lambda = \nu P S_p ~~{\rm and}~~ T_{1/2} = \frac{ln2}{\lambda},
\label{seqn10}
\vspace{0.0cm}
\end{equation}
\noindent
where $\nu$ is the assault frequency corresponding to the zero-point vibration energy $E_v$ and $S_p$ is the spectroscopic factor. 
  
\begin{figure}[t]
\vspace{0.0cm}
\eject\centerline{\epsfig{file=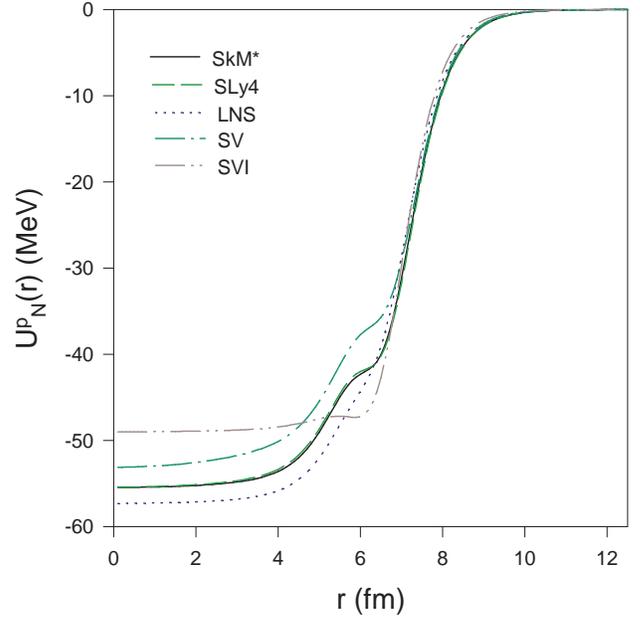,height=8.2cm,width=8.2cm}}
\caption
{ Plots of nuclear part of proton-nucleus interaction potentials for $^{185}$Bi proton emitter as functions of distance $r$ for different Skyrme sets.}
\label{fig1}
\vspace{0.0cm}
\end{figure}
\noindent
  
\section{Results and Discussions}

    The calculation of p-N nuclear potential, $U_N^p(r)$ in Eq.(8), requires the knowledge of the nucleonic density distributions in the nucleus $\rho_q(\vec{r})$, their gradients $\nabla\rho_q(\vec{r})$, $\nabla^2\rho_q(\vec{r})$ and the kinetic energy densities $\tau_q(\vec{r})$ with $q=n, p$. Since all Skyrme sets have the common feature of describing the ground state properties of nuclei over the periodic table, instead of doing Hartree-Fock (HF) calculation of a nucleus for each Skyrme set we have taken Wood-Saxon (WS) density distribution with the constraint that neutron (proton) densities are proportional to the respective neutron (proton) numbers. For the kinetic energy densities, the semi-classical approximation upto second order,

\begin{equation}
 \tau_{n(p)} = \frac{3}{5} k^2_{n(p)}\rho_{n(p)} + \frac{1}{36} \frac{(\nabla\rho_{n(p)})^2}{\rho_{n(p)}}+ \frac{1}{3}\nabla^2\rho_{n(p)},
\label{seqn11}
\vspace{0.0cm}
\end{equation}
\noindent
for neutron (proton) is used. The neutron (proton) Fermi momentum, $k_{n(p)} = [ 3\pi^2 \rho_{n(p)} ]^{1/3}$, corresponds to neutron (proton) density $\rho_{n(p)}$. The fixation of the WS parameters for the density distributions used in the present work is the same as used in Ref.\cite{BCS05}. The p-N potential thus calculated from Eq.(4) in case of $^{185}$Bi nucleus is shown in Fig.-1 as a function of distance $r$ for different Skyrme sets having wide differences in their nuclear matter properties, such as, effective mass and incompressibility. The effective mass is 0.38 for SV-set \cite{Be75}, whereas, SVI-set \cite{Be75} has value 0.95. The incompressibility value for SVI-set \cite{Be75} is 364 MeV, whereas, Sly4-set \cite{Ch98} and SKM* \cite{Ba82} have values 230 MeV and 217 MeV, respectively. The p-N potentials calculated for the Skyrme sets shown in Fig.-1 differ considerably in the interior region of the nucleus but the difference decreases towards the surface resulting in all the curves approaching the x-axis almost together in the tail region. The total potential $U^p(r)$ is obtained by adding the Coulomb and centrifugal parts to $U_N^p(r)$. The centrifugal potential is evaluated using the $l$-values for the decay processes and given in Ref.\cite{So02}. The 2nd and 3rd turning points $R_a$ and $R_b$ are obtained using the experimental $Q$-values \cite{So02} and from the relation

\begin{equation}
 U^p(R_a) = Q+E_v = U^p(R_b),
\label{seqn12}
\vspace{0.0cm}
\end{equation}
\noindent
whose solutions provide 3 turning points where $E_v  = \frac{1}{2}h\nu$ is the zero-point vibrational energy which is calculated using Eq.(5) of Ref.\cite{Po86}. Since the lifetime results are reasonably good and the number of potentials used in the present study being very large, the coefficients of the functional form for the zero-point energy are not refitted. Such a functional form facilitates the shell effects of the radioactivity to be implicitly contained in the zero point vibration energy because of its proportionality with the $Q$ value. The penetration probabilities for different proton radioactive nuclei are calculated from Eq.(9) in the cases of various Skyrme sets and half-lives obtained from Eq.(10) (for $S_p=1$) are given in Tables I-IV along with the results of the folding model potential using DDM3Y effective interaction \cite{BCS08} for comparison. Tables II-IV contains the results for the 27 Skyrme sets \cite{St03} qualifying the tests for application to neutron rich dense matter. Recently, decay properties of $^{155}$Ta are measured and the measured decay $Q$ value of 1.444(15) MeV and half life $log_{10}T(s)=-2.538^{+0.181}_{-0.207}$ \cite{Pa07} disagree with the previously reported measurements. The theoretical results of $log_{10}T(s)$ for SLy230a and DDM3Y interactions corresponding to the measured $Q$ value are $-2.13(13)$ and $-2.18(14)$, respectively. It is noteworthy that lifetimes estimated using the DDM3Y microscopic effective N-N interaction potential is as good, if not better, as those estimated by the phenomenological Skyrme interactions.

\begin{figure}[t]
\vspace{0.0cm}
\eject\centerline{\epsfig{file=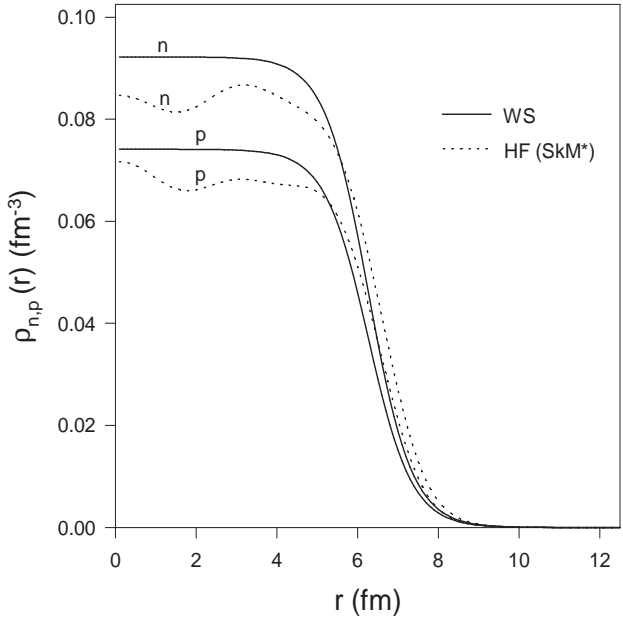,height=8.2cm,width=8.2cm}}
\caption
{ Wood-Saxon and SkM* Hartree-Fock proton and neutron density distributions as functions of distance $r$ for $^{184}$Pb nucleus (daughter of $^{185}$Bi proton emitter).}
\label{fig2}
\vspace{0.0cm}
\end{figure}
\noindent
     
\begin{figure}[t]
\vspace{0.0cm}
\eject\centerline{\epsfig{file=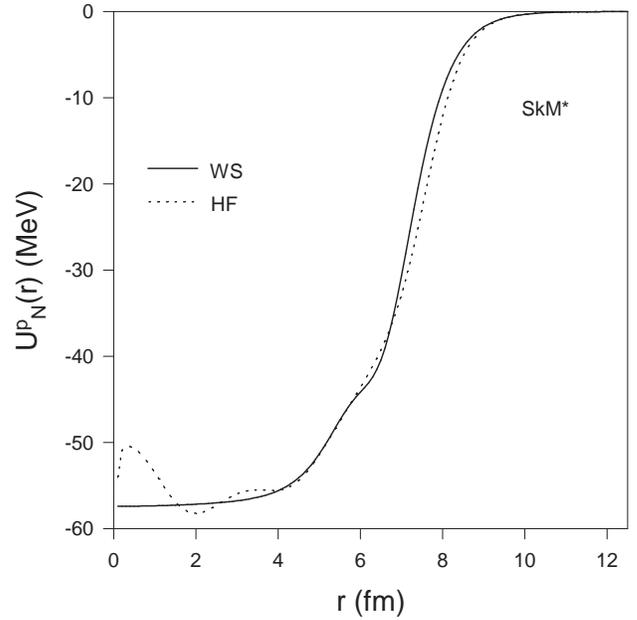,height=8.2cm,width=8.2cm}}
\caption
{ Proton-nucleus nuclear potentials calculated for $^{185}$Bi proton emitter as functions of distance $r$ for SkM* using Wood-Saxon and Hartree-Fock densities.}
\label{fig3}
\vspace{0.0cm}
\end{figure}
\noindent 
                    
    In order to examine the justification of using WS density distributions for the nuclei, we have calculated the HF densities and kinetic energy densities for neutron and proton in case of $^{185}$Bi proton emitter for the SkM* set. The p-N potential $U_N^p(r)$ is then calculated from Eq.(4) for these HF densities. The HF neutron and proton densities as well as the resulting p-N potential in case of $^{185}$Bi proton emitter for SkM* set are compared with the corresponding results obtained for WS density distribution in Fig.-2 and Fig.-3 respectively. The small differences in the p-N potential curves for the two cases (WS and HF) in the surface region in Fig.-3  do not substantially change the results of the half-lives. The results for the logarithmic half-lives are -5.40 and -5.52, respectively, for the WS and HF density distributions. The same calculations are repeated for SVI set, which has relatively high value of incompressibility, and the results are -5.28 and -5.43 for the WS and HF density distributions respectively. Since the use of self consistent HF densities does not alter the trend of the results obtained with the WS densities together with second order semi-classical kinetic energy densities, the calculations of the p-N potentials from Eq.(4) are preferred with WS densities in order to preserve the analytical simplicity. The half lives calculated using densities, its derivatives and kinetic energy densities from self consistent HF are given for all the nuclei for SkP \cite{SkP}, SkX \cite{SkX} and LNS \cite{LNS} in Table I.  

\section{Summary and Conclusion}
    
    The half-lives of the proton emitters calculated for Skyrme interaction using the energy density formalism predict similar results for the different Skyrme sets despite widely varying nuclear matter properties, particularly incompressibility and effective mass. The results of the proton radioactivity half-lives for the Skyrme interaction are also in agreement with the values obtained in case of finite range interactions, such as, JLM, DDM3Y, YENI, etc. for which the calculation of p-N potential becomes much more complicated. From the results given in Tables I-IV, it is clear that all the Skyrme sets are competent to give an overall account of the proton radioactivity half-lives. This is due to the fact that the p-N nuclear potentials are almost identical for these different Skyrme sets around the second turning point that occurs in the tail part, although in the interior region the potentials are very different for the different Skyrme parametrizations as can be seen from Fig.-1. It may be mentioned here that since the parameters of the Skyrme sets have been constrained to give binding energies and radii of nuclei over the periodic table, the variations in the nuclear matter properties of the different Skyrme sets largely manifest the change of potential in the interior region. The value of the p-N potential in the tail region is crucial in determining the barrier penetration width for the proton in the WKB method used in this work. The third turning point is being solely determined by the Coulomb potential. For the SVI set, which has relatively high value of incompressibility of 364 MeV, the location of the second turning point $R_a$ is at a distance 7.63 fm in case of $^{185}$Bi proton emitter whereas for SkM* that has incompressibility of 217 MeV, it is at 7.77 fm. The third turning point $R_b$ being the same for both the sets has the value of 65.71 fm. The barrier width varies from 58.08 fm to 57.94 fm as we go from SVI to SkM* resulting into a change in logarithmic half-life from -5.28 to -5.40. It is also clear that the Skyrme sets having relatively higher value of the incompressibility will predict relatively higher half-lives as the p-N potential approaches relatively faster to zero value causing a shift to the second turning point $R_a$ to a lower value. 

\begin{figure}[t]
\vspace{0.0cm}
\eject\centerline{\epsfig{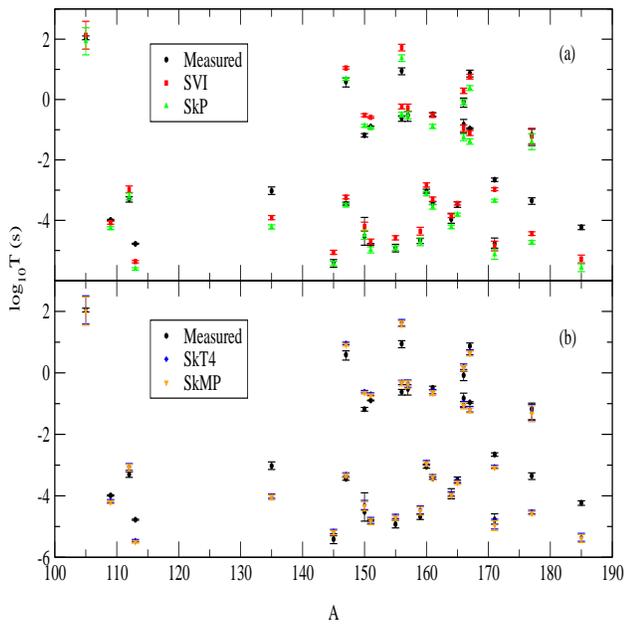}}
\caption
{ Plots of logarithmic half lives of proton emitters for (a) SVI and SkP interaction sets having extreme values of incompressibility and (b) SkT4 and  SkMP interaction sets having widely different effective masses. For details see text.}
\label{fig4}
\vspace{-0.4cm}
\end{figure}
\noindent  

    The relative importance of the nuclear matter properties on the decay half-lives have been shown in Fig.-4 (a) and (b). In examining the effect of the incompressibility, we have chosen the two Skyrme sets having almost the same value for effective mass, but widely differing in their incompressibility values. The variation of $log_{10}T(s)$ for the two such sets out of all the sets given in Tables I-IV, namely SkP and SVI, are given in Fig.-4(a). These two sets have nearly same effective masses, 1.0 and 0.95, but incompressibility varying from 201 MeV to 364 MeV. As discussed earlier, the trend that higher incompressibility predicts higher value of $log_{10}T(s)$ is evident from the Fig.-4(a). As these two sets form the two extremes of all the sets given so far as the incompressibility is concerned, the $log_{10}T(s)$ values of rest other sets remain within the results of these two sets as can also be verified from the Tables I-IV. In order to examine the effect of variation in the effective mass on the predictions of the proton emission half-lives, the results for the two Skyrme sets, namely SkT4 and SkMP which have nearly the same incompressibility values, 236 MeV and 231 MeV, but varying in their effective masses as 1.0 and 0.65 respectively, are shown in Fig.-4(b). It may be seen from this figure that in all the cases of proton emitters, that there is little dependence of the proton emission half-lives on the effective mass. Based on the findings in the framework of the present work, it can be stated that an effective interaction that accounts for the ground state bulk nuclear properties, such as, binding energies and radii of the nuclei over the periodic table can also reasonably predict the half-lives of the proton emitters. The large discrepancies in the measured values and calculated results observed in the cases of $^{113}$Cs, $^{147}$Tm, $^{150}$Lu, $^{156}$Ta, $^{156}$Ta*, $^{161}$Re*, $^{171}$Au, $^{177}$Tl and $^{185}$Bi can be attributed to the spectroscopic factor.                  
            
    The marginal sensitivity of the proton emission half-lives for different Skyrme parametrizations found in this work is in contradiction to the findings based on the results of the half-lives for the two states of $^{161}$Re obtained from quantum mechanical tunneling calculations of single particle resonance \cite{Kh07} for the Skyrme sets SkP and SkX. In their calculation the HF mean field is scaled in order to fit the $Q$ value and separated into two parts for obtaining the bound state and scattering state wave functions by solving Schrodinger equations with the two potentials. The wave functions are used to evaluate the decay width of the resonance. This method of calculating half-lives depends sensitively on the potential and wave functions. In quantum mechanical tunneling approach the HF mean field is scaled to fit the $Q$ value, whereas, in the present model assault frequency is obtained in an approximate way. Hence these important aspects, one belonging to each of the two methods of calculating the half lives need to be studied in more detail before arriving at any conclusion on the contradicting nature of the results obtained in these two methods.    

\vspace{-0.54cm}     
\section{Acknowledgments}
\vspace{-0.54cm}

    The work is supported by the UGC-DAE-CSR /KC/ 2009 / NP06 / 1354 dated 31-7-09 and covered under the SAP program of School of Physics, Sambalpur University. 

\newpage  
\begin{table*}[h]
\vspace{0.0cm}
\caption{ The results of the present calculations using the Skyrme p-N potentials are compared with the experimental values along with the results of DDM3Y \cite{BCS08}. Except $^{135}Tb$ \cite{Wo04} and $^{159}Re$ \cite{Jo06}, all the experimental $Q$ values, half lives and $l$ values are from Ref.\cite{So02}. Experimental errors in $Q$ values and corresponding errors in calculated half lives are inside parentheses. Asterisk symbol in the parent nucleus denotes isomeric state. The superscript $HF$ denotes results calculated with HF densities, its derivatives and kinetic energy densities.}
\vspace{0.0cm}
\center
\begin{tabular}{cccccccccccccc}
\hline
\hline
Parent & $l$ & $Q^{ex}$ &Measured &SII&SIII&SVI&SkM*&LNS&LNS$^{HF}$&SkP$^{HF}$&SkX$^{HF}$&DDM3Y    \\ 
$^A Z$& $\hbar$ & MeV &$log_{10}T(s)$&$log_{10}T(s)$&$log_{10}T(s)$&$log_{10}T(s)$&$log_{10}T(s)$ &$log_{10}T(s)$ &$log_{10}T(s)$&$log_{10}T(s)$&$log_{10}T(s)$&$log_{10}T(s)$ \\ 
\hline
$^{105}Sb$&2&0.491(15)&2.049$^{+0.058}_{-0.067}$&2.07(46)&2.11(46)&2.13(46)&2.00(46)&2.07(46)&2.19(46)&1.93(45)&2.11(45)&1.90(45) \\ 
$^{109}I$&2&0.829(3)&-3.987$^{+0.020}_{-0.022}$&-4.14(4)&-4.11(4)&-4.08(5)&-4.21(4)&-4.14(4)&-4.01(4)&-4.26(5)&-4.07(5)&-4.31(5) \\ 
$^{112}Cs$&2&0.824(7)&-3.301$^{+0.079}_{-0.097}$&-3.03(11)&-2.99(11)&-2.97(11)&-3.10(11)&-3.03(11)&-2.95(11)&-3.21(11)&-3.00(11)&-3.21(11) \\ 
$^{113}Cs$&2&0.978(3)&-4.777$^{+0.018}_{-0.019}$&-5.43(4)&-5.39(4)&-5.37(4)&-5.50(4)&-5.43(4)&-5.36(4)&-5.61(4)&-5.40(4)&-5.61(4) \\ 
$^{145}Tm$&5&1.753(10)&-5.409$^{+0.109}_{-0.146}$&-5.14(7)&-5.09(7)&-5.06(7)&-5.21(7)&-5.10(7)&-5.04(6)&-5.43(7)&-5.17(7)&-5.28(7) \\ 
$^{147}Tm$&5&1.071(3)&0.591$^{+0.125}_{-0.175}$&0.97(4)&1.01(4)&1.05(4)&0.89(4)&1.02(4)&1.07(4)&0.68(4)&0.94(5)&0.83(4) \\ 
$^{147}Tm^*$&2&1.139(5)&-3.444$^{+0.046}_{-0.051}$&-3.28(6)&-3.25(6)&-3.23(6)&-3.35(6)&-3.28(6)&-3.22(6)&-3.50(6)&-3.27(6)&-3.46(6) \\ 
$^{150}Lu$&5&1.283(4)&-1.180$^{+0.055}_{-0.064}$&-0.59(5)&-0.55(4)&-0.52(5)&-0.66(4)&-0.55(4)&-0.48(4)&-0.87(5)&-0.63(4)&-0.74(4) \\ 
$^{150}Lu^*$&2&1.317(15)&-4.523$^{+0.620}_{-0.301}$&-4.28(15)&-4.24(15)&-4.21(15)&-4.35(15)&-4.28(15)&-4.20(15)&-4.48(15)&-4.26(15)&-4.46(15) \\ 
$^{151}Lu$&5&1.255(3)&-0.896$^{+0.011}_{-0.012}$&-0.67(3)&-0.63(4)&-0.59(3)&-0.75(4)&-0.63(4)&-0.57(3)&-0.95(3)&-0.72(3)&-0.82(4) \\ 
$^{151}Lu^*$&2&1.332(10)&-4.796$^{+0.026}_{-0.027}$&-4.78(10)&-4.74(10)&-4.71(9)&-4.84(10)&-4.78(10)&-4.69(10)&-4.98(10)&-4.76(10)&-4.96(10) \\
$^{155}Ta$&5&1.791(10)&-4.921$^{+0.125}_{-0.125}$&-4.66(7)&-4.61(7)&-4.58(7)&-4.72(7)&-4.61(7)&-4.54(7)&-4.92(7)&-4.70(7)&-4.80(7) \\ 
$^{156}Ta$&2&1.028(5)&-0.620$^{+0.082}_{-0.101}$&-0.30(8)&-0.25(7)&-0.23(8)&-0.36(7)&-0.28(7)&-0.21(7)&-0.50(8)&-0.29(8)&-0.47(8) \\ 
$^{156}Ta^*$&5&1.130(8)&0.949$^{+0.100}_{-0.129}$&1.65(11)&1.69(10)&1.72(11)&1.58(11)&1.69(11)&1.76(10)&1.37(11)&1.60(11)&1.50(10) \\ 
$^{157}Ta$&0&0.947(7)&-0.523$^{+0.135}_{-0.198}$&-0.32(11)&-0.29(11)&-0.27(12)&-0.39(12)&-0.33(11)&-0.26(12)&-0.52(12)&-0.31(11)&-0.51(12) \\
$^{160}Re$&2&1.284(6)&-3.046$^{+0.075}_{-0.056}$&-2.90(7)&-2.86(7)&-2.83(7)&-2.96(7)&-2.89(7)&-2.82(7)&-3.12(7)&-2.90(7)&-3.08(7) \\ 
$^{161}Re$&0&1.214(6)&-3.432$^{+0.045}_{-0.049}$&-3.35(7)&-3.32(7)&-3.30(7)&-3.41(7)&-3.35(7)&-3.29(7)&-3.55(8)&-3.34(7)&-3.53(7) \\ 
$^{161}Re^*$&5&1.338(7)&-0.488$^{+0.056}_{-0.065}$&-0.60(7)&-0.55(8)&-0.52(7)&-0.67(7)&-0.55(7)&-0.51(8)&-0.89(7)&-0.65(8)&-0.75(8) \\ 
$^{164}Ir$&5&1.844(9)&-3.959$^{+0.190}_{-0.139}$&-3.92(6)&-3.88(6)&-3.85(6)&-4.00(6)&-3.89(6)&-3.83(6)&-4.22(6)&-3.98(6)&-4.08(6) \\ 
$^{165}Ir^*$&5&1.733(7)&-3.469$^{+0.082}_{-0.100}$&-3.52(5)&-3.47(5)&-3.44(5)&-3.59(5)&-3.48(5)&-3.42(6)&-3.81(5)&-3.57(5)&-3.67(5) \\
$^{166}Ir$&2&1.168(8)&-0.824$^{+0.166}_{-0.273}$&-1.02(10)&-0.97(10)&-0.95(11)&-1.09(11)&-1.00(10)&-0.96(10)&-1.25(11)&-1.01(10)&-1.19(10) \\ 
$^{166}Ir^*$&5&1.340(8)&-0.076$^{+0.125}_{-0.176}$&0.21(8)&0.25(9)&0.29(9)&0.15(9)&0.26(9)&0.30(9)&-0.09(9)&0.15(8)&0.06(9) \\ 
$^{167}Ir$&0&1.086(6)&-0.959$^{+0.024}_{-0.025}$&-1.17(9)&-1.13(9)&-1.11(8)&-1.23(9)&-1.16(9)&-1.11(9)&-1.39(9)&-1.16(9)&-1.35(8) \\ 
$^{167}Ir^*$&5&1.261(7)&0.875$^{+0.098}_{-0.127}$&0.68(8)&0.73(8)&0.76(8)&0.61(8)&0.73(8)&0.77(8)&0.38(8)&0.62(8)&0.54(8) \\ 
$^{171}Au$&0&1.469(17)&-4.770$^{+0.185}_{-0.151}$&-4.91(16)&-4.88(16)&-4.85(16)&-4.97(16)&-4.91(16)&-4.85(16)&-5.13(16)&-4.90(16)&-5.10(16) \\ 
$^{171}Au^*$&5&1.718(6)&-2.654$^{+0.054}_{-0.060}$&-3.05(5)&-2.99(4)&-2.97(5)&-3.11(5)&-3.00(4)&-2.97(5)&-3.35(5)&-3.09(5)&-3.19(5) \\ 
$^{177}Tl$&0&1.180(20)&-1.174$^{+0.191}_{-0.349}$&-1.28(26)&-1.24(26)&-1.21(26)&-1.34(26)&-1.28(26)&-1.11(27)&-1.40(26)&-1.18(26)&-1.44(26) \\ 
$^{177}Tl^*$&5&1.986(10)&-3.347$^{+0.095}_{-0.122}$&-4.52(6)&-4.47(6)&-4.44(6)&-4.58(6)&-4.48(7)&-4.34(6)&-4.73(6)&-4.47(6)&-4.64(6) \\ 
$^{185}Bi$&0&1.624(16)&-4.229$^{+0.068}_{-0.081}$&-5.34(14)&-5.30(13)&-5.28(13)&-5.40(13)&-5.34(14)&-5.29(13)&-5.57(13)&-5.35(13)&-5.53(14) \\ 
$^{135}Tb$&3&1.188(7)&-3.027$^{+0.131}_{-0.116}$&-3.98(8)&-3.94(8)&-3.91(7)&-4.05(7)&-3.98(7)&-3.93(8)&-4.22(7)&-3.98(8)&-4.16(8) \\
$^{159}Re$&5&1.816(20)&-4.678$^{+0.076}_{-0.092}$&-4.44(14)&-4.39(13)&-4.36(13)&-4.51(13)&-4.40(13)&-4.33(13)&-4.71(13)&-4.48(13)&-4.59(13) \\ \hline
\hline
\end{tabular} 
\vspace{0.0cm}
\end{table*}
\eject  

\newpage                    
\begin{table*}[h]
\vspace{0.0cm}
\caption{ The results of the present calculations using the Skyrme p-N potentials are compared with the experimental values. Except $^{135}Tb$ \cite{Wo04} and $^{159}Re$ \cite{Jo06}, all the experimental $Q$ values, half lives and $l$ values are from Ref.\cite{So02}. Experimental errors in $Q$ values and corresponding errors in calculated half lives are inside parentheses. Asterisk symbol in the parent nucleus denotes isomeric state. }
\vspace{0.0cm}
\center
\begin{tabular}{cccccccccccccc}
\hline
\hline
Parent & $l$ & $Q^{ex}$ &Measured &Gs&Rs&SGI&SV&SLy0&SLy1&SLy2&SLy3&SLy4    \\ 
$^A Z$& $\hbar$ & MeV &$log_{10}T(s)$&$log_{10}T(s)$&$log_{10}T(s)$&$log_{10}T(s)$&$log_{10}T(s)$ &$log_{10}T(s)$&$log_{10}T(s)$&$log_{10}T(s)$&$log_{10}T(s)$&$log_{10}T(s)$ \\ 
\hline
$^{105}Sb$&2&0.491(15)&2.049$^{+0.058}_{-0.067}$&2.06(46)&2.06(46)&2.05(46)&2.01(46)&1.99(46)&1.99(46)&2.00(45)&2.00(46)&2.00(45) \\ 
$^{109}I$&2&0.829(3)&-3.987$^{+0.020}_{-0.022}$&-4.15(4)&-4.15(4)&-4.17(5)&-4.20(4)&-4.22(5)&-4.23(5)&-4.22(5)&-4.22(5)&-4.22(5) \\ 
$^{112}Cs$&2&0.824(7)&-3.301$^{+0.079}_{-0.097}$&-3.04(11)&-3.04(11)&-3.06(11)&-3.10(11)&-3.11(11)&-3.11(11)&-3.11(11)&-3.11(11)&-3.11(11) \\ 
$^{113}Cs$&2&0.978(3)&-4.777$^{+0.018}_{-0.019}$&-5.44(4)&-5.44(4)&-5.46(4)&-5.49(4)&-5.52(4)&-5.52(4)&-5.51(4)&-5.51(4)&-5.51(4) \\ 
$^{145}Tm$&5&1.753(10)&-5.409$^{+0.109}_{-0.146}$&-5.11(7)&-5.11(7)&-5.15(7)&-5.21(7)&-5.23(6)&-5.24(6)&-5.23(7)&-5.23(6)&-5.23(7)\\ 
$^{147}Tm$&5&1.071(3)&0.591$^{+0.125}_{-0.175}$&1.00(4)&1.00(4)&0.96(4)&0.89(4)&0.87(4)&0.87(4)&0.87(4)&0.87(4)&0.88(4) \\ 
$^{147}Tm^*$&2&1.139(5)&-3.444$^{+0.046}_{-0.051}$&-3.29(7)&-3.29(6)&-3.31(6)&-3.35(6)&-3.37(6)&-3.37(6)&-3.36(6)&-3.37(6)&-3.37(6) \\ 
$^{150}Lu$&5&1.283(4)&-1.180$^{+0.055}_{-0.064}$&-0.57(5)&-0.57(5)&-0.61(4)&-0.67(4)&-0.69(5)&-0.70(4)&-0.68(4)&-0.69(5)&-0.68(4) \\ 
$^{150}Lu^*$&2&1.317(15)&-4.523$^{+0.620}_{-0.301}$&-4.28(15)&-4.28(15)&-4.30(15)&-4.34(15)&-4.36(15)&-4.36(15)&-4.35(15)&-4.35(16)&-4.35(15) \\ 
$^{151}Lu$&5&1.255(3)&-0.896$^{+0.011}_{-0.012}$&-0.64(3)&-0.64(3)&-0.69(4)&-0.75(3)&-0.77(3)&-0.77(4)&-0.77(3)&-0.77(3)&-0.77(4) \\ 
$^{151}Lu^*$&2&1.332(10)&-4.796$^{+0.026}_{-0.027}$&-4.77(10)&-4.77(10)&-4.80(10)&-4.84(10)&-4.86(10)&-4.86(10)&-4.86(10)&-4.86(10)&-4.86(10) \\
$^{155}Ta$&5&1.791(10)&-4.921$^{+0.125}_{-0.125}$&-4.63(6)&-4.63(7)&-4.66(7)&-4.72(7)&-4.75(7)&-4.76(7)&-4.75(7)&-4.75(7)&-4.75(7) \\ 
$^{156}Ta$&2&1.028(5)&-0.620$^{+0.082}_{-0.101}$&-0.29(7)&-0.29(7)&-0.31(7)&-0.36(8)&-0.37(7)& -0.38(7)&-0.37(8)&-0.37(7)&-0.37(8) \\ 
$^{156}Ta^*$&5&1.130(8)&0.949$^{+0.100}_{-0.129}$&1.68(10)&1.68(11)&1.64(11)&1.57(10)&1.54(11)&1.54(11)&1.56(10)&1.54(11)&1.56(10) \\ 
$^{157}Ta$&0&0.947(7)&-0.523$^{+0.135}_{-0.198}$&-0.33(12)&-0.33(12)&-0.35(12)&-0.39(12)&-0.40(11)&-0.40(11)&-0.40(11)&-0.40(11)&-0.40(11) \\
$^{160}Re$&2&1.284(6)&-3.046$^{+0.075}_{-0.056}$&-2.90(7)&-2.90(6)&-2.92(7)&-2.97(7)&-2.98(7)&-2.98(6)&-2.98(7)&-2.98(7)&-2.98(7) \\ 
$^{161}Re$&0&1.214(6)&-3.432$^{+0.045}_{-0.049}$&-3.36(7)&-3.36(7)&-3.38(7)&-3.42(7)&-3.43(7)&-3.43(7)&-3.43(7)&-3.43(7)&-3.43(7) \\ 
$^{161}Re^*$&5&1.338(7)&-0.488$^{+0.056}_{-0.065}$&-0.57(7)&-0.58(8)&-0.61(7)&-0.68(8)&-0.70(7)&-0.70(7)&-0.70(7)&-0.70(7)&-0.69(8) \\ 
$^{164}Ir$&5&1.844(9)&-3.959$^{+0.190}_{-0.139}$&-3.90(6)&-3.90(6)&-3.94(6)&-4.00(6)&-4.02(6)&-4.03(6)&-4.02(6)&-4.02(6)&-4.02(6) \\ 
$^{165}Ir^*$&5&1.733(7)&-3.469$^{+0.082}_{-0.100}$&-3.49(5)&-3.49(5)&-3.53(5)&-3.59(5)&-3.61(5)&-3.62(5)&-3.61(5)&-3.61(5)&-3.61(5) \\
$^{166}Ir$&2&1.168(8)&-0.824$^{+0.166}_{-0.273}$&-1.01(10)&-1.01(10)&-1.03(10)&-1.08(10)&-1.09(10)&-1.11(11)&-1.09(10)&-1.09(10)&-1.09(10) \\ 
$^{166}Ir^*$&5&1.340(8)&-0.076$^{+0.125}_{-0.176}$&0.24(9)&0.24(8)&0.20(9)&0.13(9)&0.11(9)&0.10(9)&0.12 (9)&0.11(9)&0.12(9) \\ 
$^{167}Ir$&0&1.086(6)&-0.959$^{+0.024}_{-0.025}$&-1.17(9)&-1.17(8)&-1.19(8)&-1.23(8)&-1.24(9)&-1.25(9)&-1.24(9)&-1.24(9)&-1.24(9) \\ 
$^{167}Ir^*$&5&1.261(7)&0.875$^{+0.098}_{-0.127}$&0.72(8)&0.72(8)&0.67(8)&0.61(8)&0.58(8)&0.58(8)&0.58(8)&0.58(8)&0.60(8) \\ 
$^{171}Au$&0&1.469(17)&-4.770$^{+0.185}_{-0.151}$&-4.90(16)&-4.91(16)&-4.93(16)&-4.98(16)&-4.99(16)&-4.99(15)&-4.99(16)&-4.99(16)&-4.99(16) \\ 
$^{171}Au^*$&5&1.718(6)&-2.654$^{+0.054}_{-0.060}$&-3.01(5)&-3.01(5)&-3.05(5)&-3.11(4)&-3.14(5)&-3.14(5)&-3.13(5)&-3.14(5)&-3.13(5) \\ 
$^{177}Tl$&0&1.180(20)&-1.174$^{+0.191}_{-0.349}$&-1.27(26)&-1.27(26)&-1.29(26)&-1.34(26)&-1.35(27)&-1.36(26)&-1.35(26)&-1.36(27)&-1.35(26) \\ 
$^{177}Tl^*$&5&1.986(10)&-3.347$^{+0.095}_{-0.122}$&-4.49(7)&-4.49(7)&-4.53(6)&-4.59(6)&-4.62(6)&-4.62(6)&-4.60(7)&-4.62(6)&-4.61(7) \\ 
$^{185}Bi$&0&1.624(16)&-4.229$^{+0.068}_{-0.081}$&-5.33(13)&-5.33(13)&-5.35(13)&-5.39(13)&-5.42(13)&-5.42(13)&-5.42(14)&-5.42(13)&-5.42(13) \\ 
$^{135}Tb$&3&1.188(7)&-3.027$^{+0.131}_{-0.116}$&-3.98(7)&-3.98(7)&-4.01(7)&-4.06(8)&-4.07(7)&-4.07(7)&-4.07(8)&-4.07(8)&-4.07(8) \\
$^{159}Re$&5&1.816(20)&-4.678$^{+0.076}_{-0.092}$&-4.41(13)&-4.41(13)&-4.45(13)&-4.51(13)&-4.53(13)&-4.53(13)&-4.53(13)&-4.53(13)&-4.53(13)\\ \hline
\hline
\end{tabular}
\vspace{0.0cm}
\end{table*}
\eject

\newpage
\begin{table*}[h]
\vspace{0.0cm}
\caption{ The results of the present calculations using the Skyrme p-N potentials are compared with the experimental values. Except $^{135}Tb$ \cite{Wo04} and $^{159}Re$ \cite{Jo06}, all the experimental $Q$ values, half lives and $l$ values are from Ref.\cite{So02}. Experimental errors in $Q$ values and corresponding errors in calculated half lives are inside parentheses. Asterisk symbol in the parent nucleus denotes isomeric state. }
\vspace{0.0cm}
\center
\begin{tabular}{cccccccccccccc}
\hline
\hline
Parent & $l$ & $Q^{ex}$ &Measured &SLy5&SLy6&SLy7&SLy8&SLy9&SLy10&SLy230a&SkI1&SkI2    \\ 
$^A Z$& $\hbar$ & MeV &$log_{10}T(s)$&$log_{10}T(s)$&$log_{10}T(s)$&$log_{10}T(s)$&$log_{10}T(s)$ &$log_{10}T(s)$&$log_{10}T(s)$&$log_{10}T(s)$&$log_{10}T(s)$&$log_{10}T(s)$ \\ 
\hline
$^{105}Sb$&2&0.491(15)&2.049$^{+0.058}_{-0.067}$&1.99(46)&2.00(46)&2.00(46)&2.00(46)&1.99(46)&2.01(46)&2.00(45)&2.06(46)&2.04(46)\\ 
$^{109}I$&2&0.829(3)&-3.987$^{+0.020}_{-0.022}$&-4.22(5)&-4.21(4)&-4.21(4)&-4.22(5)&-4.23(5)&-4.21(5)&-4.22(5)&-4.15(4)&-4.17(4)\\ 
$^{112}Cs$&2&0.824(7)&-3.301$^{+0.079}_{-0.097}$&-3.11(11)&-3.10(11)&-3.10(11)&-3.11(11)&-3.11(11)&-3.09(11)&-3.11(11)&-3.05(11)&-3.06(11)\\ 
$^{113}Cs$&2&0.978(3)&-4.777$^{+0.018}_{-0.019}$&-5.52(4)&-5.50(4)&-5.50(4)&-5.51(4)&-5.52(4)&-5.50(4)&-5.51(4)&-5.44(4)&-5.46(4)\\ 
$^{145}Tm$&5&1.753(10)&-5.409$^{+0.109}_{-0.146}$&-5.23(6)&-5.22(6)&-5.22(6)&-5.23(7)&-5.24(7)&-5.20(7)&-5.23(7)&-5.09(7)&-5.14(6)\\ 
$^{147}Tm$&5&1.071(3)&0.591$^{+0.125}_{-0.175}$&0.87(4)&0.89(4)&0.89(4)&0.87(4)&0.86(4)&0.90(4)&0.87(4)&1.03(4)&0.97(4)\\ 
$^{147}Tm^*$&2&1.139(5)&-3.444$^{+0.046}_{-0.051}$&-3.37(6)&-3.36(6)&-3.36(6)&-3.37(6)&-3.37(6)&-3.35(6)&-3.37(6)&-3.27(6)&-3.30(6)\\ 
$^{150}Lu$&5&1.283(4)&-1.180$^{+0.055}_{-0.064}$&-0.70(5)&-0.68(5)&-0.67(5)&-0.69(4)&-0.70(4)&-0.67(4)&-0.68(4)&-0.54(4)&-0.59(5)\\ 
$^{150}Lu^*$&2&1.317(15)&-4.523$^{+0.620}_{-0.301}$&-4.36(15)&-4.35(15)&-4.35(15)&-4.35(15)&-4.37(15)&-4.34(15)&-4.35(15)&-4.26(15)&-4.30(15)\\ 
$^{151}Lu$&5&1.255(3)&-0.896$^{+0.011}_{-0.012}$&-0.77(3)&-0.75(3)&-0.75(3)&-0.77(3)&-0.78(3)&-0.75(4)&-0.77(3)&-0.61(3)&-0.67(3)\\ 
$^{151}Lu^*$&2&1.332(10)&-4.796$^{+0.026}_{-0.027}$&-4.86(10)&-4.85(10)&-4.85(10)&-4.86(10)&-4.86(10)&-4.84(10)&-4.86(10)&-4.75(10)&-4.79(10)\\
$^{155}Ta$&5&1.791(10)&-4.921$^{+0.125}_{-0.125}$&-4.75(7)&-4.74(7)&-4.74(7)&-4.75(7)&-4.76(7)&-4.72(7)&-4.75(7)&-4.60(7)&-4.66(7)\\ 
$^{156}Ta$&2&1.028(5)&-0.620$^{+0.082}_{-0.101}$&-0.37(7)&-0.36(8)&-0.36(8)&-0.37(7)&-0.38(8)&-0.36(7)&-0.37(7)&-0.27(8)&-0.31(7)\\ 
$^{156}Ta^*$&5&1.130(8)&0.949$^{+0.100}_{-0.129}$&1.54(11)&1.56(11)&1.56(11)&1.56(10)&1.54(10)&1.57(11)&1.56(10)&1.71(10)&1.65(10)\\ 
$^{157}Ta$&0&0.947(7)&-0.523$^{+0.135}_{-0.198}$&-0.40(11)&-0.40(12)&-0.40(12)&-0.40(11)&-0.41(12)&-0.39(12)&-0.40(11)&-0.31(12)&-0.35(12)\\
$^{160}Re$&2&1.284(6)&-3.046$^{+0.075}_{-0.056}$&-2.98(6)&-2.97(7)&-2.97(7)&-2.98(7)&-2.98(7)&-2.96(6)&-2.98(7)&-2.88(7)&-2.91(6)\\ 
$^{161}Re$&0&1.214(6)&-3.432$^{+0.045}_{-0.049}$&-3.43(7)&-3.43(7)&-3.43(7)&-3.43(7)&-3.43(7)&-3.41(7)&-3.43(7)&-3.32(7)&-3.37(7)\\ 
$^{161}Re^*$&5&1.338(7)&-0.488$^{+0.056}_{-0.065}$&-0.70(7)&-0.68(7)&-0.68(7)&-0.70(7)&-0.71(8)&-0.67(7)&-0.70(7)&-0.53(7)&-0.60(8)\\ 
$^{164}Ir$&5&1.844(9)&-3.959$^{+0.190}_{-0.139}$&-4.02(6)&-4.00(6)&-4.00(6)&-4.02(6)&-4.03(6)&-4.00(6)&-4.02(6)&-3.87(6)&-3.93(6)\\ 
$^{165}Ir^*$&5&1.733(7)&-3.469$^{+0.082}_{-0.100}$&-3.61(5)&-3.60(6)&-3.60(5)&-3.61(5)&-3.62(5)&-3.59(5)&-3.61(5)&-3.45(5)&-3.52(5)\\
$^{166}Ir$&2&1.168(8)&-0.824$^{+0.166}_{-0.273}$&-1.09(10)&-1.09(10)&-1.09(10)&-1.09(10)&-1.11(10)&-1.08(11)&-1.09(10)&-0.99(11)&-1.03(10)\\ 
$^{166}Ir^*$&5&1.340(8)&-0.076$^{+0.125}_{-0.176}$&0.11(9)&0.13(9)&0.13(9)&0.11(9)&0.10(9)&0.13(9)&0.12(8)&0.28(8)&0.21(9)\\ 
$^{167}Ir$&0&1.086(6)&-0.959$^{+0.024}_{-0.025}$&-1.25(9)&-1.24(8)&-1.24(8)&-1.24(9)&-1.25(9)&-1.23(9)&-1.24(9)&-1.13(9)&-1.18(9)\\ 
$^{167}Ir^*$&5&1.261(7)&0.875$^{+0.098}_{-0.127}$&0.58(8)&0.60(8)&0.60(8)&0.58(8)&0.58(8)&0.61(8)&0.58(8)&0.77(8)&0.69(8)\\ 
$^{171}Au$&0&1.469(17)&-4.770$^{+0.185}_{-0.151}$&-4.99(16)&-4.99(16)&-4.99(16)&-4.99(16)&-4.99(16)&-4.97(16)&-4.99(16)&-4.88(16)&-4.92(16)\\ 
$^{171}Au^*$&5&1.718(6)&-2.654$^{+0.054}_{-0.060}$&-3.14(5)&-3.13(5)&-3.13(5)&-3.13(5)&-3.14(5)&-3.11(4)&-3.13(5)&-2.96(5)&-3.04(5)\\ 
$^{177}Tl$&0&1.180(20)&-1.174$^{+0.191}_{-0.349}$&-1.36(26)&-1.35(26)&-1.35(26)&-1.35(26)&-1.36(26)&-1.34(26)&-1.35(26)&-1.23(27)&-1.29(26)\\ 
$^{177}Tl^*$&5&1.986(10)&-3.347$^{+0.095}_{-0.122}$&-4.62(6)&-4.60(6)&-4.60(6)&-4.62(6)&-4.62(6)&-4.59(6)&-4.61(6)&-4.43(6)&-4.51(6)\\ 
$^{185}Bi$&0&1.624(16)&-4.229$^{+0.068}_{-0.081}$&-5.42(13)&-5.41(14)&-5.41(13)&-5.42(13)&-5.42(13)&-5.40(13)&-5.42(13)&-5.28(13)&-5.35(14)\\
$^{135}Tb$&3&1.188(7)&-3.027$^{+0.131}_{-0.116}$&-4.07(7)&-4.05(8)&-4.05(8)&-4.07(8)&-4.07(7)&-4.05(7)&-4.07(8)&-3.98(8)&-4.00(7) \\
$^{159}Re$&5&1.816(20)&-4.678$^{+0.076}_{-0.092}$&-4.53(13)&-4.52(13)&-4.52(13)&-4.53(13)&-4.54(13)&-4.51(13)&-4.53(13)&-4.39(13)&-4.44(13) \\ \hline
\hline
\end{tabular} 
\vspace{0.0cm}
\end{table*}
\eject

\newpage
\begin{table*}[h]
\vspace{0.0cm}
\caption{ The results of the present calculations using the Skyrme p-N potentials are compared with the experimental values. Except $^{135}Tb$ \cite{Wo04} and $^{159}Re$ \cite{Jo06}, all the experimental $Q$ values, half lives and $l$ values are from Ref.\cite{So02}. Experimental errors in $Q$ values and corresponding errors in calculated half lives are inside parentheses. Asterisk symbol in the parent nucleus denotes isomeric state. }
\vspace{0.0cm}
\center
\begin{tabular}{cccccccccccccc}
\hline
\hline
Parent & $l$ & $Q^{ex}$ &Measured &SkI3&SkI4&SkI5&SkI6&SkMP&SkO&SkO'&SkT4&SkT5    \\ 
$^A Z$& $\hbar$ & MeV &$log_{10}T(s)$&$log_{10}T(s)$&$log_{10}T(s)$&$log_{10}T(s)$&$log_{10}T(s)$ &$log_{10}T(s)$&$log_{10}T(s)$&$log_{10}T(s)$&$log_{10}T(s)$&$log_{10}T(s)$ \\ 
\hline
$^{105}Sb$&2&0.491(15)&2.049$^{+0.058}_{-0.067}$&2.03(45)&2.04(46)&2.04(46)&2.04(46)&2.01(46)&2.04(46)&2.04(46)&2.05(46)&1.99(46)\\ 
$^{109}I$&2&0.829(3)&-3.987$^{+0.020}_{-0.022}$&-4.18(5)&-4.17(4)&-4.18(4)&-4.17(4)&-4.21(5)&-4.17(5)&-4.17(4)&-4.17(5)&-4.23(5)\\ 
$^{112}Cs$&2&0.824(7)&-3.301$^{+0.079}_{-0.097}$&-3.07(11)&-3.06(11)&-3.07(11)&-3.07(11)&-3.09(11)&-3.05(11)&-3.05(11)&-3.06(11)&-3.11(11)\\ 
$^{113}Cs$&2&0.978(3)&-4.777$^{+0.018}_{-0.019}$&-5.47(4)&-5.46(4)&-5.46(4)&-5.46(4)&-5.50(4)&-5.46(4)&-5.46(4)&-5.46(4)&-5.52(4)\\ 
$^{145}Tm$&5&1.753(10)&-5.409$^{+0.109}_{-0.146}$&-5.17(6)&-5.14(7)&-5.16(7)&-5.15(6)&-5.20(7)&-5.14(7)&-5.14(7)&-5.16(7)&-5.23(7)\\ 
$^{147}Tm$&5&1.071(3)&0.591$^{+0.125}_{-0.175}$&0.94(4)&0.97(4)&0.95(4)&0.95(4)&0.91(4)&0.98(4)&0.97(4)&0.96(4)&0.87(4)\\ 
$^{147}Tm^*$&2&1.139(5)&-3.444$^{+0.046}_{-0.051}$&-3.32(6)&-3.30(6)&-3.31(6)&-3.31(6)&-3.34(6)&-3.30(6)&-3.30(6)&-3.31(6)&-3.37(6)\\ 
$^{150}Lu$&5&1.283(4)&-1.180$^{+0.055}_{-0.064}$&-0.62(4)&-0.59(5)&-0.61(5)&-0.61(4)&-0.65(4)&-0.59(4)&-0.60(5)&-0.61(4)&-0.68(4)\\ 
$^{150}Lu^*$&2&1.317(15)&-4.523$^{+0.620}_{-0.301}$&-4.32(15)&-4.30(15)&-4.30(15)&-4.30(15)&-4.34(15)&-4.28(15)&-4.30(15)&-4.30(15)&-4.35(15)\\ 
$^{151}Lu$&5&1.255(3)&-0.896$^{+0.011}_{-0.012}$&-0.70(3)&-0.67(3)&-0.69(3)&-0.69(3)&-0.73(3)&-0.66(3)&-0.67(4)&-0.69(4)&-0.77(3)\\ 
$^{151}Lu^*$&2&1.332(10)&-4.796$^{+0.026}_{-0.027}$&-4.81(10)&-4.79(10)&-4.80(10)&-4.80(10)&-4.84(10)&-4.78(10)&-4.80(10)&-4.80(10)&-4.85(10)\\
$^{155}Ta$&5&1.791(10)&-4.921$^{+0.125}_{-0.125}$&-4.69(7)&-4.66(7)&-4.67(7)&-4.67(7)&-4.72(7)&-4.65(7)&-4.66(7)&-4.67(7)&-4.74(7)\\ 
$^{156}Ta$&2&1.028(5)&-0.620$^{+0.082}_{-0.101}$&-0.33(7)&-0.31(7)&-0.32(8)&-0.31(8)&-0.34(8)&-0.30(8)&-0.31(7)&-0.31(7)&-0.37(8)\\ 
$^{156}Ta^*$&5&1.130(8)&0.949$^{+0.100}_{-0.129}$&1.62(10)&1.65(10)&1.64(10)&1.64(11)&1.59(10)&1.66(11)&1.65(10)&1.63(11)&1.56(11)\\ 
$^{157}Ta$&0&0.947(7)&-0.523$^{+0.135}_{-0.198}$&-0.37(12)&-0.35(12)&-0.35(11)&-0.35(11)&-0.38(12)&-0.33(12)&-0.35(11)&-0.35(12)&-0.40(11)\\
$^{160}Re$&2&1.284(6)&-3.046$^{+0.075}_{-0.056}$&-2.93(7)&-2.91(6)&-2.93(7)&-2.92(7)&-2.96(7)&-2.91(7)&-2.91(7)&-2.92(7)&-2.98(7)\\ 
$^{161}Re$&0&1.214(6)&-3.432$^{+0.045}_{-0.049}$&-3.38(7)&-3.37(7)&-3.38(7)&-3.38(7)&-3.41(7)&-3.36(7)&-3.37(7)&-3.38(7)&-3.43(7)\\ 
$^{161}Re^*$&5&1.338(7)&-0.488$^{+0.056}_{-0.065}$&-0.63(8)&-0.60(8)&-0.62(8)&-0.61(7)&-0.66(8)&-0.59(7)&-0.60(7)&-0.61(7)&-0.68(7)\\ 
$^{164}Ir$&5&1.844(9)&-3.959$^{+0.190}_{-0.139}$&-3.96(6)&-3.93(6)&-3.94(6)&-3.94(6)&-3.99(7)&-3.91(6)&-3.93(6)&-3.94(7)&-4.02(6)\\ 
$^{165}Ir^*$&5&1.733(7)&-3.469$^{+0.082}_{-0.100}$&-3.55(5)&-3.52(5)&-3.53(5)&-3.53(5)&-3.58(5)&-3.50(5)&-3.52(5)&-3.54(5)&-3.61(5)\\
$^{166}Ir$&2&1.168(8)&-0.824$^{+0.166}_{-0.273}$&-1.05(10)&-1.03(10)&-1.04(10)&-1.04(11)&-1.07(10)&-1.02(10)&-1.03(10)&-1.03(10)&-1.09(10)\\ 
$^{166}Ir^*$&5&1.340(8)&-0.076$^{+0.125}_{-0.176}$&0.18(9)&0.21(9)&0.19(9)&0.19(9)&0.15(9)&0.22(9)&0.21(9)&0.20(9)&0.13(9)\\ 
$^{167}Ir$&0&1.086(6)&-0.959$^{+0.024}_{-0.025}$&-1.20(9)&-1.18(9)&-1.19(9)&-1.19(9)&-1.22(8)&-1.17(9)&-1.18(8)&-1.18(9)&-1.24(9)\\ 
$^{167}Ir^*$&5&1.261(7)&0.875$^{+0.098}_{-0.127}$&0.66(8)&0.69(8)&0.67(8)&0.67(8)&0.63(8)&0.70(8)&0.69(8)&0.67(8)&0.60(8)\\ 
$^{171}Au$&0&1.469(17)&-4.770$^{+0.185}_{-0.151}$&-4.94(16)&-4.92(16)&-4.94(16)&-4.94(16)&-4.97(16)&-4.92(16)&-4.93(16)&-4.92(16)&-4.98(16)\\ 
$^{171}Au^*$&5&1.718(6)&-2.654$^{+0.054}_{-0.060}$&-3.07(5)&-3.04(4)&-3.05(4)&-3.05(5)&-3.09(4)&-3.02(4)&-3.04(5)&-3.06(5)&-3.13(5)\\ 
$^{177}Tl$&0&1.180(20)&-1.174$^{+0.191}_{-0.349}$&-1.31(26)&-1.29(26)&-1.30(26)&-1.30(26)&-1.33(26)&-1.28(26)&-1.30(26)&-1.29(26)&-1.35(26)\\ 
$^{177}Tl^*$&5&1.986(10)&-3.347$^{+0.095}_{-0.122}$&-4.54(6)&-4.51(6)&-4.53(6)&-4.53(6)&-4.57(6)&-4.50(6)&-4.51(7)&-4.53(6)&-4.60(6)\\ 
$^{185}Bi$&0&1.624(16)&-4.229$^{+0.068}_{-0.081}$&-5.37(13)&-5.35(14)&-5.35(13)&-5.35(13)&-5.39(13)&-5.33(13)&-5.35(13)&-5.35(13)&-5.40(13)\\ 
$^{135}Tb$&3&1.188(7)&-3.027$^{+0.131}_{-0.116}$&-4.02(7)&-4.00(7)&-4.02(8)&-4.02(8)&-4.05(8)&-4.00(7)&-4.00(8)&-4.01(7)&-4.07(7) \\
$^{159}Re$&5&1.816(20)&-4.678$^{+0.076}_{-0.092}$&-4.47(13)&-4.44(13)&-4.45(13)&-4.45(13)&-4.50(13)&-4.42(13)&-4.44(13)&-4.46(13)&-4.53(13) \\ \hline
\hline
\end{tabular} 
\vspace{0.0cm}
\end{table*}
\eject

\end{document}